# Structural Coupling for Microservices

Sebastiano Panichella[1][a], Mohammad Imranur Rahman[2][b] and Davide Taibi[2][c]
[1]*Zurich University of Applied Science (ZHAW), Zurich, Switzerland*
[2]*CLoWEE - Cloud and Web Engineering Group.*
*Tampere University. Tampere. 33720, Finland*
*panc@zhaw.ch, mohammadimranur.rahman@tuni.fi, davide.taibi@tuni.fi*

Abstract: Cloud-native Applications are *"distributed, elastic and horizontal-scalable systems composed of (micro)services which isolates states in a minimum of stateful components"*. Hence, an important property is to ensure a low coupling and a high cohesion among the (micro)services composing the cloud-native application.. Loosely coupled and highly cohesive services allow development teams to work in parallel, reducing the communication overhead between teams. However, despite both practitioners and researchers agreement on the importance of this general property, there are no validated metrics to effectively measure or test the actual coupling level between services. In this work, we propose ways to compute and to visualize the coupling between microservices, this by extending and adapting the concepts behind the computation of the traditional structural coupling. We validate these measures with a case study involving 17 open source projects and we provide an automatic approach to measure them. The results of this study highlight how these metrics provide to practitioners a quantitative and visual views of services compositions, which can be useful to conceive advanced systems to monitor the services evolution.

## 1 Introduction

Decomposing a monolithic system into independent services, and especially into microservices (Lewis and Fowler, 2014), is a very critical and complex task in modern applications, especially because of the lack of tools to support the decomposition of monolithic systems and the lack of clear and usable measures to evaluate the quality of the decomposed systems. Indeed, the architecture decomposition in microservices is usually performed manually and evaluated based on the human perception of software architects (Taibi et al., 2017),(Taibi et al., 2021),(Soldani et al., 2018).

A desirable property of microservices is that they should be as decoupled and as cohesive as possible (Lewis and Fowler, 2014). Specifically, while a low coupling is important in monolithic systems (Yourdon and Constantine, 1979), it is even more important in microservices, since loosely coupled services (statically and dynamically) allow the developer to make changes to their service without the need of modifying other services (Lewis and Fowler, 2014). Therefore, investigating ways to measure the evolving coupling between services is of fundamental importance, not only to increase the independence between teams, but to reduce also the level of dependability among software changes occurring in different system components. Indeed, as discussed in previous work, a high coupling can have a negative impact on reliability of changes, increasing the overall maintenance effort (since the change of one service requires to change also all the services coupled to the same service). However, besides the relevance of having a low coupling and high cohesion in microservices, there are no validated metrics to effectively measure or test the actual coupling level between the system services.

In this paper, we propose the *structural coupling* metric. An objective metrics that can be measured automatically, and that can help practitioners to understand how decoupled are their services, and eventually to reason on decoupling strategies.

We validate the *structural coupling* with a case study involving 17 open source projects, available from the "Microservice Dataset" (Rahman et al., 2019). The results of this study highlight how these metrics provide to practitioners quantitative and vi-sual views of service compositions, which can be useful to conceive advanced systems to monitor the evolution of services.

---
[a] https://orcid.org/0000-0003-4120-626X
[b] https://orcid.org/0000-0003-1430-5705
[c] https://orcid.org/0000-0002-3210-3990

**Paper structure**. The remainder of this paper is structured as follows. Section 2 describes the background and related works. Section 3 presents the proposed coupling metrics. Section 4 reports on a case study where we validate the proposed metrics while Section 5 draws conclusions of this work.

## 2 Background and Related Works

In this Section we first introduce the background and the terms adopted in this work, and then we describe the metrics we proposed to evaluate cloud-native (and microservices) based applications services composition.

### 2.1 What is a Microservice

A microservice-based system is a decentralized system that is composed of several independent small services, that communicate through different lightweight mechanisms. Commonly, microservices apply decentralized mechanisms such as choreography instead of choosing central service orchestrations. Microservice adopt domain-driven designs, that allow each microservice the responsibility for only one bounded context, providing only a limited amount of functionality serving specific business capabilities (Lewis and Fowler, 2014), (Hasselbring and Steinacker, 2017) and enabling continuous delivery (Amaral et al., 2015).

In microservices, scalability, losing coupling and high cohesion, independence, maintainability, deployment, health management and modularity are other fundamental properties of microservices (Lewis and Fowler, 2014). The development of microservices-based system require to consider several aspects of the system (Taibi et al., 2021). From the architectural point of view, developers must carefully consider the patterns adopted (Neri et al., 2020)(Taibi et al., 2018)(Taibi et al., 2019)(Taibi et al., 2020)(Taibi and Lenarduzzi, 2018). For this purpose, different tools might be adopted both before (Azadi et al., 2019) and after the migration (Pigazzini et al., 2020). Performance and complexity should be considered as well, since they are fundamental for efficient communication among microservices (Pahl and Jamshidi, 2016), (Amaral et al., 2015) and also fault handling and fault tolerance play important roles to take under control security issues (Pahl and Jamshidi, 2016), (Dragoni et al., 2017), (Martin and Panichella, 2019). Last, but not least, we need to consider that systems must be also maintained once they are deployed and therefore, developers should avoid to accumulate waste and technical debt during the development (Lenarduzzi et al., 2020)(Lenarduzzi and Taibi, 2018)(Soares de Toledo et al., 2019).

### 2.2 Cloud-Native and Microservices Metrics

Different metrics have been proposed for monolithic systems. However, several properties have been highlighted for service-based systems, and especially for microservices.

Bogner et al. (Bogner et al., 2017b) proposed a maintainability model for service-oriented systems and microservices. Engel et al. (Engel et al., 2018) proposed a set of six measures to evaluate a microservice-based system. Taibi and Systä (Taibi and Systä, 2019) proposed a decomposition framework based on process mining together with a set of metrics to evaluate the quality of the decomposition, identifying two size-related measures and a coupling measure. However, all the proposed metrics are mainly based on the manual measurement of a set of properties and they are not empirically validated. In their secondary study, Bogner et al. (Bogner et al., 2017a) highlighted that the majority of metrics explicitly designed for monolithic systems and for Service Oriented Architecture (SOA) can be also suitable to (micro)services. However, they also highlight that the different aspects of microservices can have a significant impact on the complexity of automatic metric collection, suggesting the need for specialized tool support.

We identified four groups of metrics in the literature (Table 1):

- *Service Size*. We considered six measures proposed by Engel et al. (Engel et al., 2018)and two proposed by Taibi and Systä (Taibi and Systä, 2019) with the goal of comparing two decomposition options. Moreover, we also report two metrics originally defined for SOA, that can be applied in microservices (Bogner et al., 2017a).

- *Service Complexity*. No microservice-specific measures have been proposed, but three metrics originally proposed for SOA can be applied in microservices (Bogner et al., 2017a).

- *Service Cohesion*, the degree to which the elements of a certain class belong together. It is a measure of how strongly related each piece of functionality of a software module is (Fenton,

Table 1: The Metrics Proposed in the Literature

| Group | Metric |
|---|---|
| **Service Size** | - *Number of synchronous cycles* (Engel et al., 2018)<br>- *Distribution of synchronous call per microservice* (Engel et al., 2018)<br>- *Number of synchronous dependencies* of each microservice (Engel et al., 2018)<br>- *Average size of asynchronous messages* (Engel et al., 2018)<br>- *Longest synchronous call trace* (Engel et al., 2018)<br>- *Number of classes per microservice* (Taibi and Systä, 2019)<br>- *Number of classes that need to be duplicated* (Taibi and Systä, 2019)(Taibi and Systä, 2020)<br>- *Weighted Service Interface Count (WSIC* (Hirzalla et al., 2009))*: number of exposed interface of a service be weighted on the number of parameters.<br>- *Component Balance* (Bouwers et al., 2011)(Bogner et al., 2017b)*: number and size uniformity of components (or services). Very big or very small components could be candidates for refactoring.<br>- *Number of Operations* (Shim et al., 2008)*: number of exposed interface of a service. |
| **Service Complexity** | - *Total Response for Service* (Perepletchikov et al., 2007)*: adaptation of Response for Class (RFC) (Chidamber and Kemerer, 1994) to the service level<br>- *Number of Versions concurrently used in a Service**<br>- *Service Support for Transactions** |
| **Service Cohesion** | - *Service Interface Data Cohesion (SIDC)* (Perepletchikov et al., 2007)*, the similarity of the parameters data-types between two services<br>- *Service Interface Usage Cohesion (SIUC)* (Perepletchikov et al., 2007)*:<br>($used\ operations\ per\ client/(clients \cdot operations\ in\ a\ service)$)<br>- *Total Service Interface Cohesion* (Perepletchikov et al., 2007): average between SIDC and SIUC |
| **Service Coupling** | - *Coupling Between Microservices (CBM)* (Taibi and Systä, 2019). Extension of the CBO, ratio between the number of calls to other services and the number of classes of the microservice<br>- *Absolute Importance of the Service (AIS)* (Rud et al., 2006)(Bogner et al., 2017b)* number of clients that invoke at least one operation to the service.<br>- *Absolute Dependence of the Service (ADS)* (Rud et al., 2006)* number of other services that a service depends on<br>- *Absolute Criticality of the Service* (Rud et al., 2006)* defined as: ACS(S) = AIS(S) × ADS(S)<br>- *Services Interdependence in the System (SIY)* (Rud et al., 2006)(Bogner et al., 2017b)*: Number of service pairs bidirectionally dependent on each other. If such dependencies between microservices exist, services could be merged. |

*Metrics Adopted in SOA, that could be suitable for microservices (Bogner et al., 2017a)

1991). High cohesion makes the reasoning easy and limits the dependencies (Kramer and Kaindl, 2004). No specific measures have been defined for cloud-native systems or for SOA. Bogner et al.(Bogner et al., 2017a) propose to use two cohesion metrics in microservices-based systems

- *Service Coupling*, the degree or indication of the strength of interdependence and interconnections of a service with other services. Two metrics have been proposed for measuring microservice cohesion (Taibi and Systä, 2019)(Bogner et al., 2017b). Moreover, Bogner et al. (Bogner et al., 2017a) also proposed to use four object-oriented and SOA specific metric proposed by (Rud et al., 2006) in the context of microservices.

Our works extends and complement the proposed coupling metrics for coupling by proposing clear measurement procedures, and a tool to automatically detect them in microservice, an an approach to visualize them.

# 3 The Proposed Coupling Metrics for Microservices

In this section, we introduce the Structural Coupling, a metric that can be used to measure the coupling among microservices. In particular, such metrics are inspired by previous work (Savic et al., 2017) from the software engineering research field, which proposed different ways to measure coupling among software artifacts, this to study software evolution dynamics of complex software systems.

Thus, we first describe the concepts behind the definition of these traditional coupling metrics, describing how they complement each other. Then we provide the formal definition of the metric we defined for studying the evolution, complexity, and relations among microservices.

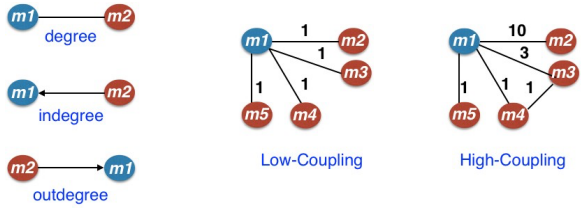

Figure 1: Structural Coupling: graph representation

## 3.1 Traditional Definitions of Structural Coupling

The concept of *high cohesion and low coupling* is one of the basic design principles in the software engineering research field (Yourdon and Constantine, 1979). According to such principles, the *coupling between modules* of a software system has to be as minimal as possible, and, at the same time keeping strong relations between software artifacts composing the individual modules.

**Structural Coupling**: A software module (or artifact) *A* is structurally coupled with another module *B*, if code/structural dependencies exist among them (Savic et al., 2017). Higher is the number of dependencies among these modules, higher is the level of coupling. The concept of structural coupling can be easily represented as a directed or indirect *collaboration graph* (Myers, 2003) (see Figure 1), in which the nodes correspond to the system modules and the weighted edges represent the code dependencies among these nodes. A low structural coupling is important to allow changes in an individual module without propagating them in other modules (high maintainability). A high structural coupling led to bugs and changes propagating among modules of systems (low maintainability).

## 3.2 Microservice Coupling Measures

Referring to the conceptual definition of structural coupling ((Savic et al., 2017)), a service *A* is structurally coupled with another service *B*, if code/structural dependencies exist among artifacts composing them. Thus, higher is the number of dependencies among these services, higher is the level of coupling. A low structural coupling is important to allow changes in an individual service without propagating them to other modules/artifacts of other services (high maintainability). A high structural coupling led to bugs and changes propagating among modules of different services (low maintainability), as well as low developer productivity (e.g., developers have to co-ordinate their development work involving different services). More formally, we computed the structural coupling among a service *s1* and service *s2* as reported in the following formula (inspired by the *Definition 10* in page 13 of the work by Savic *et al.* (Savic et al., 2017)):

$$StructuralCoupling(s1, s2) = 1 - \frac{1}{(degree(s1,s2))} \cdot LWF * GWF \quad (1)$$

In this definition of structural coupling, dependencies among two services *s1* and *s2* are weighted considering both Local Weighting Factor (LWF), considering the degree and in-degree of s1 with s2, and the Global Weighting Factor (GWF), considering the max degree among all services of the system, weighting factors:

$$LocalWeightFactor(s1, s2) = \frac{1 + outdegree(s1,s2)}{1 + degree(s1,s2)}$$

$$GlobalWeightFactor(s1, s2) = \frac{degree(s1,s2)}{max(degree(all\_services))} \quad (2)$$

This re-weighted structural coupling measurement ensures that the actual coupling value between *s1* and *s2* range between [0-1] and that these values depend also on the general dependencies distributed to other services.

- ***degree(s1,s2)*** is the number of all structural dependencies between s1 and s2;
- the ***outdegree(s1,s2)*** is the actual number of static dependencies, among the total one, that are directed from s1 to s2; ***in-degree(s1,s2)*** is the actual number of static dependencies, among the total one, that are directed from s2 to s1;
- the ***max(degree(all_services))*** corresponds to the max number of dependencies (i.e., max degree) among all (possible pairs of) services of the system.

## 3.3 Example

Starting from the system depicted in Figure 2, we now describe how to calculate the structural coupling. The system adopted as example is composed by five microservices, connected together directly.

Table 2 reports an example of metrics proposed in the literature, calculated for the example-system depicted in Figure 2. In particular, we represent the size of each microservices (number of classes), the in-degree of each microservice (the number of incoming service calls), the out-degree (the number of outgoing calls), and the degree (sum of in-degree and out-degree).

Table 3 shows the example of the calculation Local Weight Factor (LWF), Global Weight Factor (GWF) and Structural Coupling (SC) on the same system.

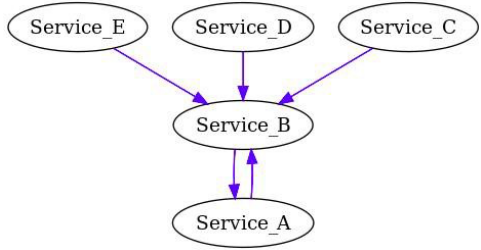

Figure 2: The Example of Microservices-based System

Table 2: Example of metrics for the system in Figure 2

|   | in-degree | out-degree | degree | #classes |
|---|---|---|---|---|
| A | 4 | 1 | 5 | 50 |
| B | 0 | 1 | 1 | 10 |
| C | 0 | 1 | 1 | 11 |
| D | 0 | 1 | 1 | 17 |
| E | 0 | 1 | 2 | 30 |

## 4 Validation

**Case and Subjects Selection**: We selected the 17 projects developed in Java and using Docker from the Microservice Dataset (Rahman et al., 2019). Table 4 describes the 17 selected projects, providing information on the number of microservices in each system (#MS), the size of each system in lines of code (#KLOC), the number of commits and the number of dependencies (#Dep) .

**Data Collection Procedure**: For each project, we counted the number of classes per microservice, to compare our measure with the measures proposed in the literature. Then, we calculated the coupling metrics proposed in the literature, together with the Structural Coupling proposed in Section 3.

We generated a csv file collecting metrics in-degree, out-degree, degree, #classes (number of classes), LOC (Lines Of Code). For each metric SIY, LWF, GWF and SC we generated a matrix to calculate the dependency between all couples of microservices.

**Analysis procedure** For each system, we draw a chart reporting SC for each pairs of MS. This graphical representation is useful to understand which are the MS in the system with the highest coupling. In order to compare the different measures, we also compute descriptive statistics of SC for each microservice.

The results of the analysis are available in the replication package[1]

### 4.1 Results

From Table 5, we can already observe that the average values of structural coupling metrics, computed among all services of each project, tend to be different than CBM and that the simple degree metric. This already confirms our initial conjecture that such metric can provide a different view of the services composition of microservice-based applications.

To facilitate the interpretation of such metrics, and to further confirm our conjecture, we provide a graph representation of some of the projects (not all of them for reason of space) in Table 4. In particular, as discussed in Section 3, given the computed couplings values among the services it is possible to represent the service composition of a project as a directed graph, in which the nodes correspond to the microservices and the weighted edges represent the coupling dependencies among these nodes. Thus, by leveraging the R packages *igraph*,*sna*, and *ggplot2*, we generate the graph that is possible to observe in Figure 3.

To simplify the interpretation of the various generated graph (Figure 3) we used different coloring strategies: (i) we colored in **green** all the nodes (services) that are a hub in the services coupling network (i.e., have a high degree); (i) in **yellow** we colored all the nodes (services) that act as bridge between two or more services in the services coupling network; (i) we colored in **blue** all the nodes (services) that have a high out-degree, compared to their in-degree. The remaining nodes are colored in **red**. Finally, the size of nodes (i.e., services) in the graphs reflect the degree of each service in the services coupling network (e.g., higher is the relative degree of a service compared to other services, bigger will be its size ).

In the case of Spring (Table 4), the microservices are connected to each other via an API Gateway, and this is reflected in the structural coupling: no service is no structurally coupled with other services. In Figure 3, we can see that the three coupling metrics provide different view of the service compositions. In this specific case, the structural coupling shows a very well structured service composition. Indeed, in a few cases we observe a node with a high number of structural dependencies with other nodes.

---
[1]Replication Package https://github.com/clowee/Structural-Coupling-for-Microservices

Table 3: SIY, LWF, GWF and SC for the system in Figure 2

|   | LWF | | | | | GWF | | | | | SC | | | | |
|---|---|---|---|---|---|---|---|---|---|---|---|---|---|---|---|
|   | A | B | C | D | E | A | B | C | D | E | A | B | C | D | E |
| A | 0 | 0.66 | 1 | 1 | 1 | 0 | 2 | 0 | 0 | 0 | | | | | |
| B | 0 | 0 | 0.5 | 0.5 | 0.5 | 0 | 0 | 1 | 1 | 1 | 0.33 | | | 0.5 | 0.5 |
| C | 1 | 0 | 0 | 1 | 1 | 0 | 0 | 0 | 0 | 0 | | 0.5 | | | |
| D | | 1 | 0 | 1 | 0 | 1 | 0 | 0 | 0 | 0 | 0 | | | | |

Table 4: The selected projects

| Project Name | #Ms. | KLOC | #Commits | #Dep. |
|---|---|---|---|---|
| CQRS microservice application | 7 | 1.632 | 86 | 3 |
| E-Commerce App | 7 | 0.967 | 20 | 4 |
| EnterprisePlanner | 5 | 4.264 | 49 | 2 |
| eShopOnContainers | 25 | 69.874 | 3246 | 18 |
| FTGO - Restaurant Management | 13 | 9.366 | 172 | 9 |
| Lakeside Mutual Insurance | 8 | 19.363 | 12 | 7 |
| Microservice Blog post | 9 | 1.536 | 90 | 7 |
| Microservices book | 6 | 2.417 | 127 | 5 |
| Open-loyalty | 5 | 16.641 | 71 | 2 |
| Pitstop - Garage Management | 13 | 34.625 | 198 | 9 |
| Robot Shop | 12 | 2.523 | 208 | 8 |
| Share bike (Chinese) | 9 | 3.02 | 62 | 6 |
| Spinnaker | 10 | 33.822 | 1669 | 6 |
| Spring Cloud Microservice | 10 | 2.333 | 35 | 9 |
| Spring PetClinic | 8 | 2.475 | 658 | 7 |
| Spring-cloud-netflix | 9 | 0.419 | 61 | 6 |
| Vehicle tracking | 8 | 5.462 | 116 | 5 |

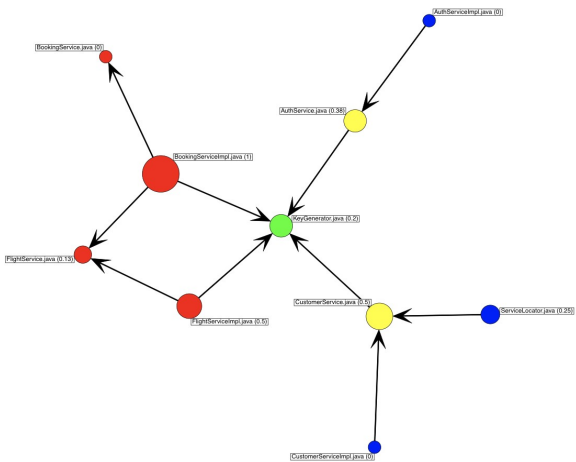

Figure 3: Example of Structural Coupling Graph representation

Our findings show that the usage of structural coupling can be of high relevance for developers interested to observe how their services are decomposed. Moreover, these metrics provide a rather different, but also complementary views of the services decomposition. We believe that such metrics can be used for guiding refactoring operations or re-modularization at the level of microservice composition. In addition, they can be used for making quick diagnosis on potential bad developers coordination practices, when evolving/migrating applications to the cloud.

It is interesting to note that CBM (Taibi and Systä, 2019) can not be computed in six projects (out of 17), while Structural Coupling can anyways be applied. It is also important to note that our proposal, together with the graphical representation of the structural coupling, allows to easily see microservices with a high out-degree and to graphically compare the Structural Coupling of each node.

## 5 Discussions and Conclusions

Practitioners and researchers agree that microservices must be lowly coupled and highly cohesive. The development of microservice-based systems is growing, however at the best of our knowledge, there are no validated metrics to evaluate coupling and cohesion between services. Some researchers (Bogner et al., 2017a) proposed to extend coupling measures adopted for SOA but these measures have never been validated nor used in the microservice domain.

In order to help practitioners to clearly identify coupling between services, in this work, we introduced the Structural Coupling, a metric based on the structural dependencies between services.

We validated the structural coupling measure on 17 Open Source projects developed with a microservice architectural pattern and we proposed a visualization to graphically represent the measure. Results show that structural coupling easily shows the degree of coupling between existing services, and the visualization provided can be adopted to easily spot coupling issues in (micro)services. Differently than other coupling metrics for microservices, structural coupling seems to be always applicable, while in some case, CBM (Taibi and Systä, 2019) is not applicable, since its denominator can be zero.

Future works include the packaging of script to calculate the measures and to generate the visualizations into an Open Source tool. As for the application of the structural coupling to different systems, other type of connections between services should be investigated. As an example, microservices might be connected using publisher-subscriber mechanisms, or using other REST principles such as HATEOAS (Hypermedia as the Engine of Application State) that enable loose coupling by design. We are planning to validate these metrics on industrial case studies, to an-alyze the perceived usefulness (e.g., by analyzing user and developers feedback (Panichella et al., 2015)) of the visualization and to analyze further correlations between coupling and maintenance effort or other software qualities perceived by the developers. Future works also include the definition and validation of metrics to evaluate the system decomposition, including cohesion metrics. Moreover, future works include the application of Structural Coupling to other cloud-native technologies such as serverless functions.

Table 5: Results of the Coupling Metrics Applied to the 17 projects.

| Project Name | Degree | | | | SC | | | | | CBM | | | | |
|---|---|---|---|---|---|---|---|---|---|---|---|---|---|---|
| | Max | Avg | Med. | Stdev | Tot | Max | Avg | Med. | Stdev | Tot | Max | Avg | Med. | Stdev |
| CQRS microservice appl. | 1 | 1 | 1 | 0.0 | 7.25 | 0.88 | 0.80 | 0.75 | 0.06 | 2.5 | 1.0 | 0.35 | 1.0 | 0.23 |
| E-Commerce App | 2 | 1.14 | 1.0 | 0.34 | 8.87 | 0.88 | 0.80 | 0.75 | 0.06 | 3.27 | 1.0 | 0.46 | 1.0 | 0.42 |
| EnterprisePlanner | 3 | 1.00 | 1.00 | 0.00 | 3.50 | 0.83 | 0.70 | 0.67 | 0.07 | | | | | |
| eShopOnContainers | 8 | 1.16 | 1.00 | 0.46 | 71.06 | 0.94 | 0.91 | 0.94 | 0.03 | | | | | |
| FTGO - Restaurant Man. | 2 | 1.13 | 1.0 | 0.33 | 24.00 | 0.9 | 0.86 | 0.9 | 0.05 | 0.39 | 0.12 | 0.03 | 0.04 | 0.34 |
| Lakeside Mutual Ins. | 3 | 1.67 | 1.00 | 1.05 | 6.67 | 0.83 | 0.74 | 0.67 | 0.08 | 1.12 | 1.00 | 0.12 | 0.03 | 0.39 |
| Microservice Blog post | 4 | 1.10 | 1.00 | 0.30 | 13.75 | 0.88 | 0.81 | 0.75 | 0.06 | 3.61 | 1.00 | 0.36 | 0.50 | 0.25 |
| Microservices book | 1 | 1.83 | 1.00 | 1.86 | 1.00 | 0.50 | 0.20 | 0.00 | 0.24 | 2.75 | 1.00 | 0.46 | 0.30 | 0.41 |
| Open-loyalty | 3 | 1.20 | 1.00 | 0.40 | 2.83 | 0.83 | 0.71 | 0.67 | 0.07 | | | | | |
| Pitstop - Garage Manag. | 3 | 1.15 | 1.00 | 0.53 | 14.50 | 0.83 | 0.76 | 0.83 | 0.08 | 1.14 | 0.33 | 0.09 | 0.08 | 0.11 |
| Robot Shop | 4 | 1.50 | 1.00 | 0.76 | 9.38 | 0.88 | 0.78 | 0.75 | 0.05 | | | | | |
| Share bike (Chinese) | 3 | 1.10 | 1.00 | 0.30 | 11.33 | 0.83 | 0.76 | 0.83 | 0.08 | 2.38 | 1.00 | 0.24 | 0.13 | 0.43 |
| Spinnaker | 7 | 1.20 | 1.00 | 0.60 | 18.79 | 0.93 | 0.89 | 0.93 | 0.04 | | | | | |
| Spring Cloud Micros. | 7 | 1.10 | 1.00 | 0.30 | 23.21 | 0.93 | 0.89 | 0.89 | 0.04 | 5.75 | 1.00 | 0.57 | 1.00 | 0.41 |
| Spring PetClinic | 2 | 1.09 | 1.00 | 0.29 | 7.75 | 0.75 | 0.60 | 0.50 | 0.12 | 3.01 | 1.00 | 0.27 | 0.35 | 0.29 |
| Spring-cloud-netflix | 7 | 1.11 | 1.00 | 0.31 | 23.29 | 0.93 | 0.90 | 0.93 | 0.04 | 5.75 | 1.00 | 0.64 | 1.00 | 0.29 |
| Vehicle tracking | 4 | 1.00 | 1.00 | 0.00 | 13.88 | 0.88 | 0.82 | 0.88 | 0.06 | | | | | |

Table 6: Results of the LWF and GWF of the 17 projects.

| Project Name | LWF | | | | | GWF | | | | |
|---|---|---|---|---|---|---|---|---|---|---|
| | Tot | Max | Avg | Med. | Stdev | Tot | Max | Avg | Med. | Stdev |
| CQRS microservice application | 7.0 | 1.0 | 0.78 | 1.0 | 0.25 | 2.25 | 0.25 | 0.25 | 0.25 | 0.0 |
| E-Commerce App | 8.5 | 1.0 | 0.77 | 1.0 | 0.25 | 2.75 | 0.25 | 0.25 | 0.25 | 0.0 |
| EnterprisePlanner | 4.5 | 1.0 | 0.9 | 1.0 | 0.2 | 1.67 | 0.33 | 0.33 | 0.33 | 0.0 |
| eShopOnContainers | 55.5 | 1.0 | 0.71 | 0.5 | 0.25 | 9.75 | 0.13 | 0.13 | 0.13 | 0.0 |
| FTGO - Restaurant Management | 20.0 | 1.0 | 0.71 | 0.5 | 0.25 | 5.6 | 0.2 | 0.2 | 0.2 | 0.0 |
| Lakeside Mutual Insurance | 7.0 | 1.0 | 0.78 | 1.0 | 0.25 | 3.0 | 0.33 | 0.33 | 0.33 | 0.0 |
| Microservice Blog post | 13.0 | 1.0 | 0.76 | 1.0 | 0.25 | 4.25 | 0.25 | 0.25 | 0.25 | 0.0 |
| Microservices book | 4.0 | 1.0 | 0.8 | 1.0 | 0.24 | 5.0 | 1.0 | 1.0 | 1.0 | 0.0 |
| Open-loyalty | 3.5 | 1.0 | 0.88 | 1.0 | 0.22 | 1.33 | 0.33 | 0.33 | 0.33 | 0.0 |
| Pitstop - Garage Management | 13.5 | 1.0 | 0.71 | 0.5 | 0.25 | 6.33 | 0.33 | 0.33 | 0.33 | 0.0 |
| Robot Shop | 10.5 | 1.0 | 0.88 | 1.0 | 0.22 | 3.0 | 0.25 | 0.25 | 0.25 | 0.0 |
| Share bike (Chinese) | 11.0 | 1.0 | 0.73 | 0.5 | 0.25 | 5.0 | 0.33 | 0.33 | 0.33 | 0.0 |
| Spinnaker | 15.5 | 1.0 | 0.74 | 0.5 | 0.25 | 3.0 | 0.14 | 0.14 | 0.14 | 0.0 |
| Spring Cloud Microservice | 19.5 | 1.0 | 0.75 | 0.75 | 0.25 | 3.71 | 0.14 | 0.14 | 0.14 | 0.0 |
| Spring PetClinic | 10.5 | 1.0 | 0.81 | 1.0 | 0.24 | 6.5 | 0.5 | 0.5 | 0.5 | 0.0 |
| Spring-cloud-netflix | 19.0 | 1.0 | 0.73 | 0.5 | 0.25 | 3.71 | 0.14 | 0.14 | 0.14 | 0.0 |
| Vehicle tracking | 12.5 | 1.0 | 0.74 | 0.5 | 0.25 | 4.25 | 0.25 | 0.25 | 0.25 | 0.0 |